\documentclass[12 pt]{amsart}
\usepackage{url}
\usepackage{graphicx}
\usepackage[normalem]{ulem}
\usepackage{rotating}
\usepackage[margin=1in]{geometry}
\usepackage{setspace}
\usepackage{booktabs}
\usepackage{ctable}
\usepackage{exscale}
\usepackage{amsbsy}
\usepackage{amsfonts}
\usepackage{amssymb}
\usepackage{amsmath}
\usepackage{amsaddr}
\usepackage{natbib}
\usepackage{color}
\newlength{\savearraycolsep}

\newcommand{\bds}[1]{\boldsymbol{#1}}
\newcommand{\bse}{\boldsymbol{\eta}}

\newcommand{\bsG}{\boldsymbol{G}}
\newcommand{\bsR}{\boldsymbol{R}}

\newcommand{\bsS}{\boldsymbol{S}}

\newcommand{\bsX}{\boldsymbol{X}}

\newcommand{\bss}{\boldsymbol{s}}
\newcommand{\bsz}{\boldsymbol{z}}
\newcommand{\bsx}{\boldsymbol{x}}
\newcommand{\bsw}{\boldsymbol{w}}
\newcommand{\bsy}{\boldsymbol{y}}

\newcommand{\bsr}{\boldsymbol{r}}
\newcommand{\bstheta}{\boldsymbol{\theta}}

\newcommand{\moparm}{\boldsymbol{\Psi}}

\newcommand{\bsbeta}{\boldsymbol{\beta}}
\newcommand{\bsGamma}{\boldsymbol{\Gamma}}
\newcommand{\te}[1]{\textrm{#1}}

\newcommand{\ud}{\mathrm{d}}

\author[Karl, Yang, and Lohr]{Andrew T. Karl}
\address{Adsurgo LLC}
\author{Yan Yang}
\address{Arizona State University}
\author{Sharon L. Lohr}
\address{Westat}

\title[Correlated Random Effects Model for Nonignorable Missing Data]{A Correlated Random Effects Model for Nonignorable Missing Data in Value-Added Assessment of Teacher Effects}
\date{}

\begin{document}
\begin{abstract}
Value-added models have been widely used to assess the contributions of individual teachers and schools to students' academic growth based on longitudinal student achievement outcomes.  There is concern, however, that ignoring the presence of missing values, which are common in longitudinal studies, can bias  teachers' value-added scores. In this article, a flexible correlated random effects model is developed that  jointly models the
student responses and the student missing data indicators. Both the student responses and the missing data mechanism depend on latent teacher effects as well as latent student effects, and the correlation
between the sets of random effects adjusts teachers' value-added scores for informative missing data. The methods are illustrated with data from calculus classes at a large public university and with data from an elementary school district.
\end{abstract}

\maketitle
\section*{NOTICE}
This is the author's version of a work that was accepted for publication in the \textit{Journal of Educational and Behavioral Statistics}. Changes resulting from the publishing process, such as peer review, editing, corrections, structural formatting, and other quality control mechanisms may not be reflected in this document. Changes may have been made to this work since it was submitted for publication. A definitive version was subsequently published in \textit{Journal of Educational and Behavioral Statistics}, [VOL38, (2013)] DOI:10.3102/1076998613494819


\section{INTRODUCTION}

With increased focus on accountability in education has come increased interest in measuring teacher and school contributions toward their students' learning. Assessing teachers  solely by their current-year students' scores on a standardized test is widely recognized to penalize teachers of disadvantaged students  \citep{braun2}; the measures of teacher effectiveness are biased because teacher effects are confounded with their students' characteristics.  Value-added models (VAMs) attempt to reduce this bias  by estimating the effects teachers have on the academic growth of their students. Rather than simply calculating the average test score for a classroom, as might be done in a naive performance analysis, VAMs control for information on the students' backgrounds, the students' individual test score histories, and contributions of previous teachers to the students' learning. The simplest VAMs use a gain score as a response \citep{Hanushek1971} or include the previous year's test score as a covariate in a regression model \citep{Rowan}; these control for the student's background through the previous year's test score and possibly other covariates. The Colorado Growth Model \citep{betebenner} uses the previous year's test score as a covariate in a quantile regression model.
Other VAMs have been defined using mixed models
\citep{sanders97,raudenbush02,ballou,mc04,lock07,harris10,wright,mariano10}, in which the response is a vector of student scores over time and teacher contributions are modeled through random effects. In mixed models, the empirical best linear unbiased predictors (EBLUPs) of the teacher random effects serve as the VAM scores.  As noted by \citet{lock07}, these EBLUPs summarize unexplained heterogeneity at the classroom level, though they are often referred to as ``teacher effects.''

VAM scores may be used for a variety of purposes, from identifying needs for professional development to high-stakes purposes such as promoting or firing teachers or closing schools.
Many researchers and policymakers have expressed concern about whether  VAM scores have sufficient accuracy
for high-stakes purposes \citep{nas,Baker2010,braun,briggs,Harris2011}.
If the model assumptions are met and the model contains all relevant information, the VAM scores from that model will be unbiased estimates of the teacher effects on the responses measured. The model assumptions are strong, however, and there is concern about how often some of the assumptions are met in practice. The models assume that students are assigned to teachers randomly or in a non-informative manner \citep{Rothstein2010}, that the responses are valid measures of student achievement \citep{koretz}, and that all relevant information is captured in the model. \citet{Lohr2012} discusses the assumptions of various models and shows how violations of the assumptions may be used to manipulate VAM scores.

The models cited above also assume that every student has complete data over the time period studied, or
that missing data patterns have no information about teacher effectiveness.
\citet{ballou} note that longitudinal mixed model approaches allow  students to have missing test scores for some years by including a partial vector of responses, but such analyses assume that the data are missing at random.
Missing data are ubiquitous in longitudinal education data. Students drop courses, change schools, move away, or may be absent on the day of a test. Inference based on analyses of data where some observations are missing requires assumptions about the nature of the missing data. In the college setting, students in calculus 2 who do not finish calculus 3 will have missing data for calculus 3. The missingness may be relevant to estimates of the calculus 2 teachers' contributions. A student who is poorly prepared for calculus 3 may drop the class despite having received a high grade in calculus 2. Or, in the elementary- or secondary-school settings, it is possible that low-performing students might be discouraged from taking a standardized exam \citep{ryan, nytnew}. In a simplistic example, suppose that students are randomized to one of several classrooms and
a gain score model is used. If a teacher were to discourage her weakest students from taking the exam, she could inflate her class average and thus her ranking.

The assumptions about missing data made by VAMs have been recognized as a potential problem for their use in teacher evaluation \citep{mc03, braun2,Amrein2008}. \citet{mc05} and \citet{wright04} explore the impact of the presence of missing data on VAMs, though they do not perform a joint analysis of the test scores and missing data indicators. To date, the only thorough investigation of the impact of missing data on VAMs by jointly modeling the test scores and missing data process comes from \citet{mc10}.  They use selection and pattern-mixture models for the missing data indicators with Bayesian inference, attributing attendance to intrinsic student -- but not teacher -- characteristics.

In this paper, we develop a new multiple response, multiple membership mixed model that allows the missing data mechanism to depend on teachers as well as students.
This model allows detection of
teachers' possible effects on their students' future course taking or their students' attendance during an exam.
Because the true responses are missing, the models cannot be used to say for certain that teacher VAM scores would change if the missing data were taken into account, but the model in this paper allows exploration of possible effects of missing data on the teacher rankings through a sensitivity analysis \citep{xu}. If  the rankings of teacher effects change depending on the assumptions made about the structure of the missing data mechanism, then the possible dependence on missing data should be considered when contemplating high-stakes usages of VAM scores.  Even if the teacher effects do not show sensitivity to the structure of the missing data mechanism, the model may be useful as a diagnostic tool. In some situations, no relationship would be expected between the teacher effects and the corresponding effects in the missing data mechanism. By fitting the model and examining a scatter plot of the effects, unusual cases may be discovered.

The paper is organized as follows.
Section \ref{sec:model} of this paper presents background on missing data analyses and the framework for modeling the test scores and the missing data mechanism jointly. Section \ref{sec:application} applies the joint model to calculus data from a large public university.
Structures available within the model are used to perform a sensitivity analysis on the teacher rankings produced when analyzing a data set containing semester calculus grades.
Section \ref{sec:gradeschool} summarizes the results of the model when applied to elementary school math scores.
Finally, Section~\ref{sec:summary} discusses implications of model estimates for uses of VAMs and other applications in which the models developed in this paper can describe potential effects of missing data.

\section{A Correlated Random Effects Model}\label{sec:model}
Let $y_{ig}$ be the potential response (often, a test score) of student $i$ at time $g$,  for $g=1,\ldots,T$, with $\bds{y_i}=(y_{i1},\ldots,y_{iT})^{\prime}$ and $\bds{y}=(\bds{y}_1^{\prime},\ldots,\bds{y}_n^{\prime})^{\prime}$.  The indicator variable
\begin{displaymath}
r_{ig} = \left\{ \begin{array}{ll}
1 & \te{if $y_{ig}$ is observed}\\
0 & \te{otherwise}
\end{array} \right.
\end{displaymath}
tracks whether the planned measurement on student $i$ at time $g$ is observed or missing. Let $\bds{r}_i=(r_{i1},\ldots,r_{iT})^{\prime}$ and $\bds{r}=(\bds{r}_1^{\prime},\ldots,\bds{r}_n^{\prime})^{\prime}$. The complete data vector $\bds{y}=\{\bds{y}^o,\bds{y}^m\}$ consists of both the observed data $\bds{y}^o$ and the missing data $\bds{y}^m$. The vector $\bds{y}^o$ consists of the values $y_{ig}$ such that $r_{ig}=1$, and $\bds{y}^m$ consists of the values $y_{ig}$ that would have been observed if the observations were not missing.
Since $r_{ig}=1$ if we observe the value $y_{ig}$, we refer to the model generating the $r_{ig}$ as the attendance process, where by ``attendance'' in a particular year we simply mean that a student has a test score recorded for that year.
We refer to the model generating the scores $y_{ig}$ as the longitudinal or the score process.

Data may be missing from a study for several reasons, and the cause of the missingness determines the degree to which the missing data affect the analysis. If data are missing completely at random, then the joint likelihood of the longitudinal and attendance processes factors cleanly, and there is no need for joint modeling, since the longitudinal and attendance processes are independent. Likewise, if the data are missing at random (MAR) and the parameters for the longitudinal and missingness processes are distinct, then the missing data mechanism is said to be ignorable for likelihood inference \citep{little87}. However, if the missing data are missing not at random (MNAR) and hence nonignorable, then the longitudinal and missingness processes
cannot be factored in the likelihood; they must be modeled jointly to explore the effects of missingness on estimates in the longitudinal process.

\citet{mc10} have developed selection and pattern-mixture VAMs for nonignorable missing data in which the missing data mechanism depends on latent effects of the students. We expand the availability of VAMs for data with potentially nonignorable missing data by presenting a correlated-parameter model (CPM), a generalization of a shared-parameter model (SPM: \citealp{wu88}). In the CPM, random effects are included for the latent teacher and student effects in the longitudinal model, a different set of random effects are included for the latent teacher and student effects in the attendance model, and the two sets of random effects are allowed to be correlated \citep{lin09}. Allowing correlated rather than shared random effects as in the SPM avoids the SPM's restriction that the random effects have the same variance and structure.
The CPM proposed in this paper allows the missing data mechanism to depend on the effects of teachers as well as students. This gives more flexibility in detecting sensitivity to missing data, since it is plausible that the missing data trajectory of students could depend on their current and former teachers.

The CPM produces the observed data likelihood via the factorization
\begin{equation}\label{eq:cpmlik}
f(\bds{y^o,r})=\iint{f(\bsy^o|\bse_{score})f(\bsr|\bse_{attnd})f(\bse_{score},\bse_{attnd})\ud \bse_{score}\ud \bse_{attnd}}
\end{equation}
where $f(\bse_{score},\bse_{attnd})$ is the density of a multivariate normal distribution. The vector $\bse_{score}$ contains random student and teacher intercepts for the longitudinal process, while the vector $\bse_{attnd}$ contains a flexible combination of student and/or teacher effects for the attendance mechanism. The CPM assumes that the longitudinal and attendance processes are conditionally independent, given the random effects.

CPMs make different assumptions on the joint model than selection and pattern-mixture models (e.g. conditional independence) and present an alternative approach for missing data modeling. The CPM framework allows for straightforward inclusion of teacher history in the modeling of the dropout mechanism. The EBLUPs of the classroom effects in the attendance model provide a direct method of evaluating the frequency with which teachers' students have missing data. Since the attendance model estimates the probability that a given observation would be recorded, a larger EBLUP for a classroom effect in the attendance model indicates that students who took that particular class are more likely to complete the next year than students who took another class that year (i.e. with another teacher). It would, however, be unrealistic to expect the effect of a teacher on student learning to be identical to the effect of the teacher on the future student attendance, so $\bse_{score}$ and $\bse_{attnd}$ are assumed correlated rather than identical.

\subsection{The Observed Data Model}\label{ssec:gp}
We now present the model $f(\bsy^o|\bse_{score})$ for student scores $\bds{y}^o$ using information about the history of observations on each student and each student's teacher-history. We use the Generalized Persistence (GP) model of \citet{mariano10} for the longitudinal mechanism.
The GP model is among the most general of the mixed models used for VAMs, and contains many of the other mixed models as special cases. If the data are MAR, the model in (\ref{eq:cpmlik}) reduces to the GP model. Suppose a data set tracks a cohort of  $n$ students over $T$ years. The GP model assumes a linear mixed model as follows:
\begin{equation}\label{eq:model}
   y^o_{ig}=\bsx_{ig}^{\prime}\bsbeta_{score}+\bss_{ig}^{\prime}\bse_{score}+\epsilon_{ig}
\end{equation}
where $y^o_{ig}$ denotes the score for student $i$ during year $g$, for $i=1,\ldots,n$, and $g\in A_i$; $A_i$ is the set of years in which student $i$ is observed. Students are taught by one of $m_g$ teachers in each year $g$. We will also refer to the vector of concatenated student scores, $\bds{y^o}=(\bds{y^o_1}^{\prime},\ldots,\bds{y^o_{n}}^{\prime})^{\prime}$, where $\bds{y^o_i}=(y^o_{ig})$. The matrix $\bsX$, with rows $\bds{x}_{ig}^{\prime}$, is the design matrix for the vector $\bsbeta_{score}$ of student and teacher level covariates such as demographic information or years of teaching experience. The matrix $\bsS$, with rows $\bds{s}_{ig}^{\prime}$, indicates which students and teachers are associated with the responses in $\bsy^o$.

The random effects vector $\bse_{score}=\left[\bds{\delta}^{\prime}_{score}\; \bstheta^{\prime}_{score}\right]^{\prime}$ has two components. Student $i$ has a latent effect $\delta_i$ that represents an underlying level of achievement not explained by the fixed covariates, and $\bds{\delta}^{\prime}_{score}=\left(\delta_1,\ldots,\delta_n\right)^{\prime}$ . We assume that $\delta_1,\ldots,\delta_n$ are independent and identically distributed $N_1(0,\Gamma_{stu})$ random variables. This represents a slight departure from \citet{mariano10}, who model the intra-student correlation in an unstructured error covariance matrix. However, that structure is not as amenable to the joint model for attendance because it precludes the possibility of including student effects in the attendance model. As a result, we model the intra-student correlation with random effects, similar to the VAM used by \citet{mc10}. When the responses $y_{ig}$ all have the same scale for $g=1, \ldots, T$, this leads to a compound-symmetry covariance structure for the students. If the student random effects are omitted from the missing data mechanism, the intra-student correlation may be modeled in an unstructured error covariance matrix as done by \citet{mariano10}.

The GP model estimates the effect of teachers on students in the year that they teach them, their lasting effect on the next year's score, and so on. Following the notation of \citet{mariano10}, we let $\theta_{g[jt]}$ represent the effect for the $j$-th grade-$g$ teacher on a student's grade $t$ score. A grade $g=1,\ldots,T$ teacher has $K_g=T-g+1$ effects. Thus $\bds{\theta}_{g[j\cdot]}$ gives the vector of current and future year effects of the $j$-th grade $g$ teacher. The vector $\bstheta_{score}$ concatenates the $\bds{\theta}_{g[j\cdot]}$ effects for all grades and teachers. The model is able to distinguish between the persistence effect of former teachers and the current effect of the present teacher because the students are not nested at the teacher level. The design matrix $\bsS$ of the random effects has rows $s^{\prime}_{ig}$, and may be partitioned into two blocks $\bsS=\left[\bsS_1\;\bsS_2\right]$. $\bsS_1$ contains a 1 in column $i$ if the observation is for student $i$, and $\bsS_2$ contains $1$'s in entries corresponding to teachers who could affect that response. We specify the structure and distribution of the random effects in Section \ref{ssec:joint}.

The error terms are distributed as $\bds{\epsilon} \sim N(\bds{0},\bds{R})$ where $\bds{R}$ is a diagonal matrix with entries coming from the set $\left\{\sigma^2_1,\ldots,\sigma^2_T\right\}$, depending on the year of the observation. In addition, we assume $\te{cov}(\bds{\eta}_{score},\bds{\epsilon})=\bds{0}$.

\subsection{The Attendance  Model}
In the attendance model, the probability $p_{ig}$ that student $i$ provides a score at time $g$ (i.e., $r_{ig} = 1$) depends on covariates and latent teacher and/or student effects. We use a threshold model for $p_{ig}$, the conditional probability that $r_{ig}=1$ \citep{mcculloch94}. Using a probit link, the generalized linear mixed model is
\begin{eqnarray*}
r_{ig}|\bds{\eta}_{attnd}&\sim&\te{Bin}(1,p_{ig})\\
\Phi^{-1}(p_{ig})&=&\bsw^{\prime}_{ig}\bsbeta_{attnd}+\bsz^{\prime}_{ig}\bds{\eta}_{attnd}
\end{eqnarray*}

The vectors $\bsw^{\prime}_{ig}$ and $\bsz^{\prime}_{ig}$ describe which fixed and random effects are thought to be related to the response mechanism. The vector of fixed effects $\bds{\beta}_{attnd}$ of the attendance model will be different from the $\bds{\beta}_{score}$ of the observed model. It will represent a baseline propensity for attendance at each level of the fixed effects. Furthermore, the attendance model requires that there is at least one missing observation at each level of each categorical fixed effect in the attendance mechanism. Otherwise, the data suffer from quasi-complete separation \citep{allison}. In that case, the maximum likelihood estimate for the particular fixed effect does not exist.

We may include either random teacher effects, random student effects, or both in $\bse_{attnd}$. The structure of the random effects is flexible, and may be modified depending on the goals of the study. This flexibility provides the means for performing a sensitivity analysis. When jointly modeling MNAR data, the CPM makes untestable assumptions about the nature of the relationship between the observed data and attendance processes. \citet{molen08} show that it is not possible to perform an overall test of MNAR versus MAR since every MNAR model has an MAR counterpart that provides the same fit to the observed data but different predictions for the unobserved data. The plausibility of the assumed model cannot be tested empirically, and as a result it is necessary to fit several alternatives of the attendance model to check the sensitivity of the inference to the choice of joint modeling structure \citep{xu}.

The student effects in the attendance model, if included, will be denoted by $\delta^{attnd}_i$. The teacher effects in the attendance model will be denoted by $\Lambda_{g[j]}$. These effects may be structured in a number of different ways. In our application in Section \ref{sec:application}, $\Lambda_{g[j]}$ represents the effect that the $j$-th grade $g$ teacher has on the probability of his or her students being measured in year $g+1$. This effect measures how likely it is that students are observed in the year after studying under a particular teacher. This effect is not calculated for teachers in the last year of observations (year $T$) because no information is available on the future dropout patterns of students of those teachers. This feature of the model would detect instructors whose students drop out (of the school or sequence of courses) at a relatively high rate. We refer to these effects as the ``attendance effects'' of the grade $g$ teachers, since they measure the rate with which students complete year $g+1$.
This models the effects of teachers on their students' future course-taking as well as on their completion of subsequent courses.

In other settings, it makes more sense to model the effect of missing data in the current year, $g$.  For example, in the grade-school application in Section \ref{sec:gradeschool}, we structure the missing data mechanism to measure the proportion of each grade-$g$ teacher's students who actually take the standardized exam in that year.	In the calculus example, we may wish to distinguish between students who drop out of a calculus 3 course and those who never enrolled. If information about students who drop courses is available, it would be reasonable to use the attendance effect of a grade-$g$ teacher to model the proportion of students who complete their course. The model is flexible and allows for many variations on the implementation of the missing data mechanism. The attendance mechanism may be used to model the effects of year $g$ teachers on attendance in year $g$, on attendance in year $g+1$, or on both, assuming different random effects for the two years. When both teacher and student effects are included in the attendance model, it is important to make sure those effects are defined to model the same concept.

The conditional density of $r_{ig}$ given the random effects vector $\bse_{attnd}$ (which contains the effects $\delta^{attnd}_i$ and $\Lambda_{g[j]}$)  is\begin{align*}
f(r_{ig}|\bds{\eta}_{attnd})&=\Phi\left(\left(-1\right)^{1-r_{ig}}\left[\bsw_{ig}^{\prime}\bsbeta_{attnd}+\bsz_{ig}^{\prime}\bds{\eta}_{attnd}\right]\right).
\end{align*}
As with the $y_{ig}$, we assume the $r_{ig}$ are conditionally independent given the random effects, yielding
\begin{align*}
f(\bds{r}|\bds{\eta}_{attnd})=\prod_{i=1}^{n}\prod_{g=1}^{T}\Phi\left(\left(-1\right)^{1-r_{ig}}\left[\bsw_{ig}^{\prime}\bsbeta_{attnd}+\bsz_{ig}^{\prime}\bds{\eta}_{attnd}\right]\right).
\end{align*}

\subsection{The Joint Model}\label{ssec:joint}

In typical usage, VAMs assume that missing data are MAR. Inference is intended to be on $\bsy=(\bsy^o,\bsy^m)$, but only the $\bsy^o$ have been observed. When data are MNAR, $f(\bsy^o)$ is not the correct likelihood to maximize because $\bsr$ provides information about the distribution of $\bsy$. As a result, the longitudinal and attendance processes must be modeled jointly and $f(\bsy^o,\bsr)$ must be maximized. We construct the joint model via the correlated random effects factorization (\ref{eq:cpmlik}).

We concatenate the random effects vectors $\bse_{score}$ and $\bse_{attnd}$ into a single random effects vector, $\bse$. To ensure that the $\te{cov}(\bse)=\bsG$ matrix is block-diagonal, we structure the $\bds{\eta}$ vector as
\begin{multline}\label{eq:eta}
\bds{\eta}=\left(\delta_1,\delta_1^{{attnd}},\ldots,\delta_n,\delta_n^{{attnd}},\bds{\theta_{1[1\cdot]}},{\Lambda_{1[1]}},\ldots,\bds{\theta_{1[m_1\cdot]}},{\Lambda_{1[m_1]}},\bds{\theta_{2[1\cdot]}},{\Lambda_{2[1]}},\ldots,\right.\\
\left.\bds{\theta_{2[m_2\cdot]}},{\Lambda_{2[m_2]}},\ldots, \bds{\theta_{T[m_T\cdot]}}\right)^{\prime}
\end{multline}
We model the random student effects and their counterparts for the attendance mechanism, if they are included, as $\left(\delta_i,\delta^{attnd}_i\right)^{\prime}\sim N_2\left(\bds{0},\bsGamma_{\te{stu}}\right)$
where $\bsGamma_{\te{stu}}$ is a $2\times 2$ unstructured covariance matrix. If the random student effects are not included in the attendance  model, simply omit the $\delta^{attnd}_i$ from $\bse$ and model $\delta_i\sim N_1\left(0,\Gamma_{\te{stu}}\right)$. The teacher effects are assumed independent of the student effects and distributed as
\begin{displaymath}
\left\{ \begin{array}{ll}
\left(\bstheta^{\prime}_{g[j\cdot]},{\Lambda}^{\prime}_{g[j]}\right)^{\prime}\sim N_{K_g+1}\left(\bds{0},\Gamma_g\right) & \textrm{if } g\neq T\\
\left(\bstheta^{\prime}_{g[j\cdot]}\right)^{\prime}\sim N_{K_g}\left(\bds{0},\Gamma_g\right) & \textrm{if } g=T
\end{array} \right.
\end{displaymath}
where $\Gamma_g$ is an unstructured covariance matrix.
Then
\begin{equation}\label{eq:G}
\bds{G}=\te{cov}(\bse)=\te{blockdiag}\left(\bsGamma_{\te{stu}},\ldots,\bsGamma_{\te{stu}},\bds{\Gamma}_1,\ldots,\bds{\Gamma}_1,\ldots,\bds{\Gamma}_T,\ldots,\bds{\Gamma}_T\right)\end{equation}
where there are $n$ copies of $\bsGamma_{\te{stu}}$, and for each $g=1,\ldots,T$ there are $m_g$ copies of $\bsGamma_g$ in $\bsG$. The $\bsR$ matrix for $f(\bsy^o|\bse_{score})$ is unchanged from Section \ref{ssec:gp}. The log-likelihood for the joint model (\ref{eq:cpmlik}) may be expressed as
\begin{align}
\label{eq:loglike}
l(\moparm)&=\log\iint\ \prod_{i=1}^{n}\left\{ \prod_{g\in A_i}f(y^o_{ig}|\bse_{score})\prod_{g=1}^{T} f(r_{ig}|\bse_{attnd})\right\}f(\bse_{score},\bse_{attnd})\ud \bse_{score} \ud\bse_{attnd}
\end{align}
where
\begin{align*}
f(y^o_{ig}|\bse_{score})&\propto\left(\sigma_g^2\right)^{-1/2}\te{exp}\left[{-\left(y^o_{ig}-\bsx^{\prime}_{ig}\bsbeta_{score}-\bss^{\prime}_{ig}\bse_{score}\right)^{2}}/({2\sigma_g^2}) \right],\\
f(r_{ig}|\bse_{attnd})&=\Phi\left[\left(-1\right)^{1-r_{ig}}\left(\bsw^{\prime}_{ig}\bsbeta_{attnd}+\bsz^{\prime}_{ig}\bse_{attnd}\right)\right],\\
f(\bse_{score},\bse_{attnd})&=f(\bds{\eta})\propto \te{det}\left(\bsG\right)^{-1/2}\te{exp} \left[-({\bse^{\prime}\bsG^{-1}\bse)}/{2}\right],
\end{align*}
$A_i$ is the set of years in which student $i$ has an observation, and $\moparm$ is a vector of the model parameters.

Note that the models are specified separately: the model of the test scores $y_{ig}$ contains only the parameters $\bsbeta_{score}$ and the random effects $\delta_i$ and $\bse_{score}$; the model of the attendance indicators $r_{ig}$ contains only the parameters $\bsbeta_{attnd}$ and the random effects $\delta_i^{attnd}$ and $\bse_{attnd}$. The effects $\bse_{score}$ and $\bse_{attnd}$ are related through the correlation structure in the matrix $\bsG$. If student $i$ is absent at time $g$, there will be no observation for $y_{ig}$, but $r_{ig}=0$ will still be modeled: the correlation between the random effects in the two models means that the missing value contributes to the estimates of student and teacher effects in the test score model.

\subsection{Estimation}

The joint model presents a high-dimensional integration problem when calculating the marginal distribution of the observed data in (\ref{eq:loglike}). The source of the problem is twofold, due to the presence of a nonlinear link in the integrand for the modeling of the binary attendance process and the multiple membership structure of VAMs. The random effects' correlation structure is not nested, which means that the integral over the random effects cannot be factored into a product of low-dimensional integrals (e.g. one- or two-dimensional integrals). Even under the assumption of MAR and without the integration problem, the GP model is computationally demanding because of its random effects structure. \citet{mariano10} notice sensitivity to the choice of prior distributions for the covariance matrices when estimating the GP model with Bayesian methods. \citet{karlem} use an EM algorithm to develop an efficient maximum likelihood routine for estimating the GP model \citep{mariano10} under an assumption of MAR. The EM algorithm is available through the R \citep{R} package GPvam \citep{gpvam}.
The general method for estimating the parameters of non-nested, multiple-response GLMMs developed in \citet{karlcgs} is used to perform calculations for the CPM in this paper. This method makes use of first-order and fully exponential Laplace approximations for the intractable integrals that appear in the E step \citep{steele, riz09}.

\section{Effects of Missing Data in Calculus Classes}\label{sec:application}

This section applies the model to data on calculus grades from a large public university. \citet{broatch10} use a subset of these data in their analyses. The data set tracks 3557 students who took calculus 2 and possibly calculus 3 at the university. A total of 184 calculus 2 classes are included from Fall 2000 through Spring 2005. In addition, 144 calculus 3 classes from Spring of 2001 through Spring of 2006 are included. Students who took only calculus 3 during the study are omitted.  Each classroom is treated as a separate effect. Effects corresponding to different classes taught by the same teacher are assumed to be independent. An alternative model could be fit in which classes taught by the same
teacher are nested within that teacher and an additional random effect added at the teacher
level. In that case, it would be expected that the mean responses of classes taught by the 
same teacher would be positively correlated. Accounting for this correlation would result in slightly
larger standard errors for the estimated teacher effects. Another approach would be to introduce additional parameters to the appropriate off-block-diagonal components of $\bsG$, explicitly modeling the correlation between classroom effects belonging to the same teacher.

Analysis focuses on the grades assigned to students, which are converted to the corresponding value on a four-point scale. The scores in the data set are collectively centered and standardized. With +/- grades, there are eight possible numeric values for the student scores. The normal approximation for the error terms seems reasonable, though the quality of the approximation would deteriorate as the number of distinct grades decreases.

\subsection{Sensitivity Analysis}
In this data set, only 2140 of the 3557 students who completed calculus 2 also completed calculus 3. Longitudinal sequences in the university setting often have a different pattern of missing data than longitudinal data sets in the elementary school setting, because 
missing data in universities are often due to students' decisions to drop out of college, to change majors, or simply not to complete the calculus sequence. These decisions may be influenced by the students' previous or current instructors. In the models shown here, the attendance variable for calculus 3 is modeled as a function of the effect of the calculus 2 instructor. Some students may have such a poor experience with a particular instructor that they decide to not to take the next course in the sequence, or upon beginning the next course find themselves unprepared and drop out. Of course, a student's completion of calculus 3 is a function of many other things besides his experience with his calculus 2 instructor.

Our goal is not to select a particular attendance mechanism, but rather to test the sensitivity of teacher EBLUPs to assumptions about missing observations. As \citet{kenward} discuss, focusing attention on one particular MNAR model is no better than ignoring MNAR models. The observed data cannot provide evidence for or against the MAR assumption without an \textit{a priori} assumption about the correct form of the MNAR model \citep{Rhoads}. The choice of attendance mechanism must be made from a subject-matter perspective. When an alternate attendance mechanism provides a plausible representation of the missing data process and yields substantially different teacher effects from the test score model, then the accuracy of the MAR rankings is questionable. While the lack of sensitivity of the EBLUPs to different choices of attendance mechanisms strengthens our confidence in the results, it is always possible that the missing observations are nonignorable according to an untested attendance mechanism.

We fit a model using just the yearly means as fixed effects in both the score and attendance models (Model 1), as well as a model which includes gender, race/ethnicity, and SAT quantitative score as covariates in both the score and attendance models (Model 2). Because some of the students do not take the SAT, we treat the SAT quantitative score (SATQ) as a categorical variable with six categories: the five quintiles of scores, with a sixth category for students who did not take the SAT. Because the student scores come from non-standardized class grades, the current year teacher effects reflect the tendency of individual teachers to assign above- or below-average grades, and not necessarily the effectiveness of their teaching. The future year effects of calculus 2 teachers, however, reflect how well each teacher's former students performed in comparison to their new calculus 3 classmates. 
Our investigation focuses on these future year effects.

While not every student who takes calculus 2 does so with the intention of taking calculus 3, we may expect to see, on average, a certain proportion of calculus 2 students going on to complete calculus 3. In this example, we construct the attendance mechanism to measure the proportion of students from calculus 2 classes who complete calculus 3. To perform a sensitivity analysis, we fit an MAR model and compare its estimated teacher effects to those from three different MNAR models.

In the model we will call MNAR-t, we include a random teacher effect for calculus 2 teachers in the attendance mechanism that is correlated with the corresponding teacher effects from the observed data mechanism and measures the proportion of each teacher's students who go on to complete calculus 3. The model MNAR-s models calculus 3 completion as a function of student random effects. Even though only one binary observation is made on each student, we are able to fit this model because the predicted student effects in the attendance mechanism borrow strength from their correlation with the student effects from the observed data mechanism. Finally, MNAR-b contains both random student and teacher effects in the attendance mechanism. The appropriate attendance process cannot be chosen by empirical investigation of the observed data (including examination of the log-likelihood) since the observed data do not provide information to support one particular MNAR model over another \citep{fitz,xu}. Instead, we compare the estimated teacher effects across different models, looking for sensitivity to the assumptions about the nature of the missing data.

\subsection{Results}

The parameter estimates for Model 1 appear in Table~\ref{tab:ASU}. The covariance parameter estimates for Model 2 are very similar. The estimates for the fixed effects of Model 2 appear in Table~\ref{tab:model1}. The yearly means in the observed data model are represented by $\mu_i^y$, for $i=1,2$. The value $\mu_2^r$ gives the estimated proportion, e.g. $\Phi(0.246)=0.597$, of calculus 2 students who complete calculus 3. The other parameters follow the same notation as used in Section~\ref{sec:model}. Also listed for each model are -2 times the Laplace approximated log-likelihood ($-2l$) and the correlation ($\rho$) of the predicted calculus 2 future year effects with those from the MAR model. This correlation provides a summary of the sensitivity of the teacher rankings to assumptions about the nature of student dropout under different models for the attendance mechanism. Using selection and pattern mixture models to model the dropout process as a function of student effects, \citet{mc10} found values of $\rho$ that were all greater than 0.97. MNAR-s provides the analog of their models using correlated random effects, and yields $\rho=0.994$. Likewise, MNAR-b does not produce teacher effects that are substantially different from the MAR model. However, MNAR-t reorders the teacher effects, producing $\rho=0.881$.

\begin{figure}
\caption{Calculus 2 Future Year Effects: MAR vs. MNAR-t. The solid circle represents a teachers whose VAM score changes substantially under different assumptions for the missing data mechanism.}
\label{plot:asu_joint_mar_2}
\centering
\includegraphics[scale=.5]{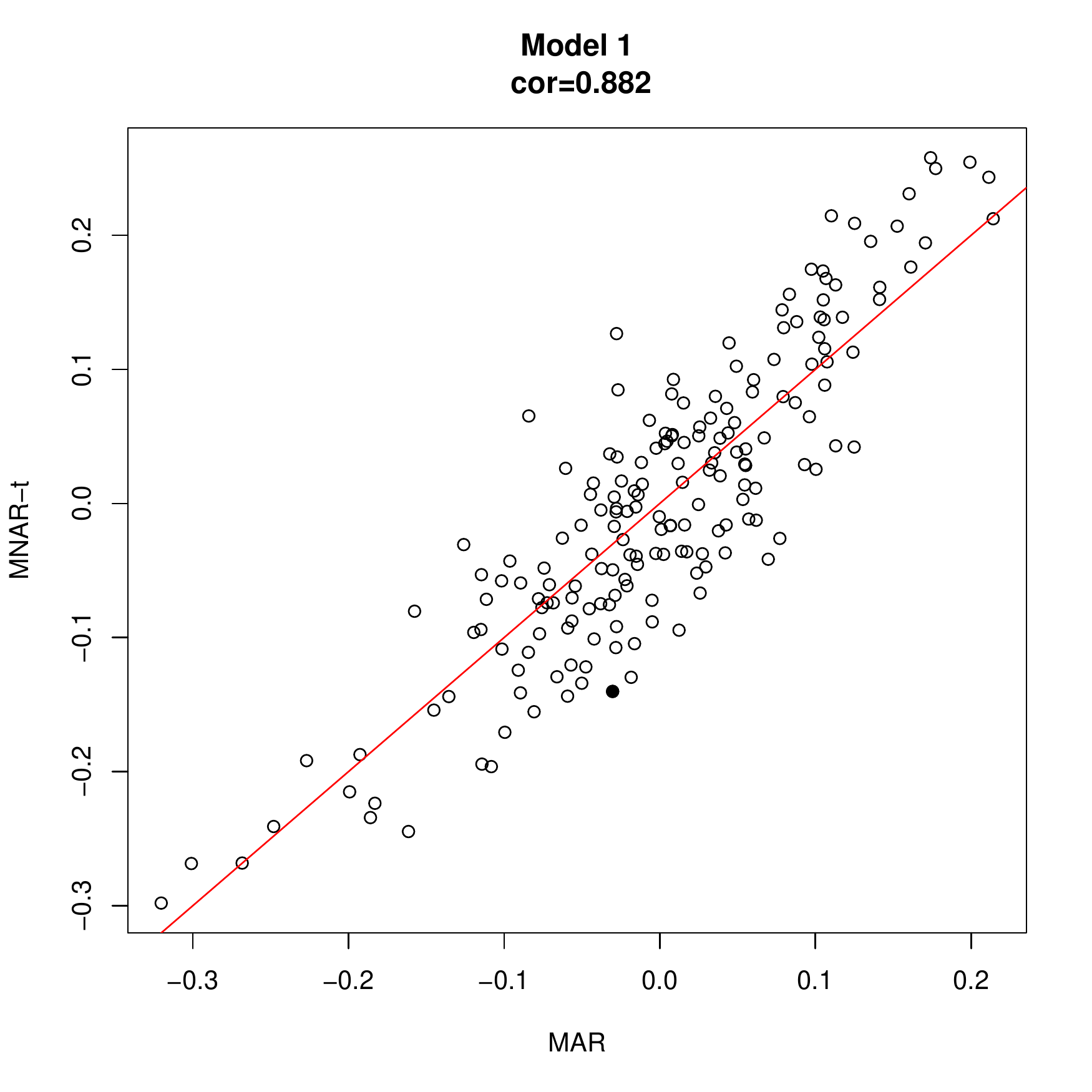}
\includegraphics[scale=.5]{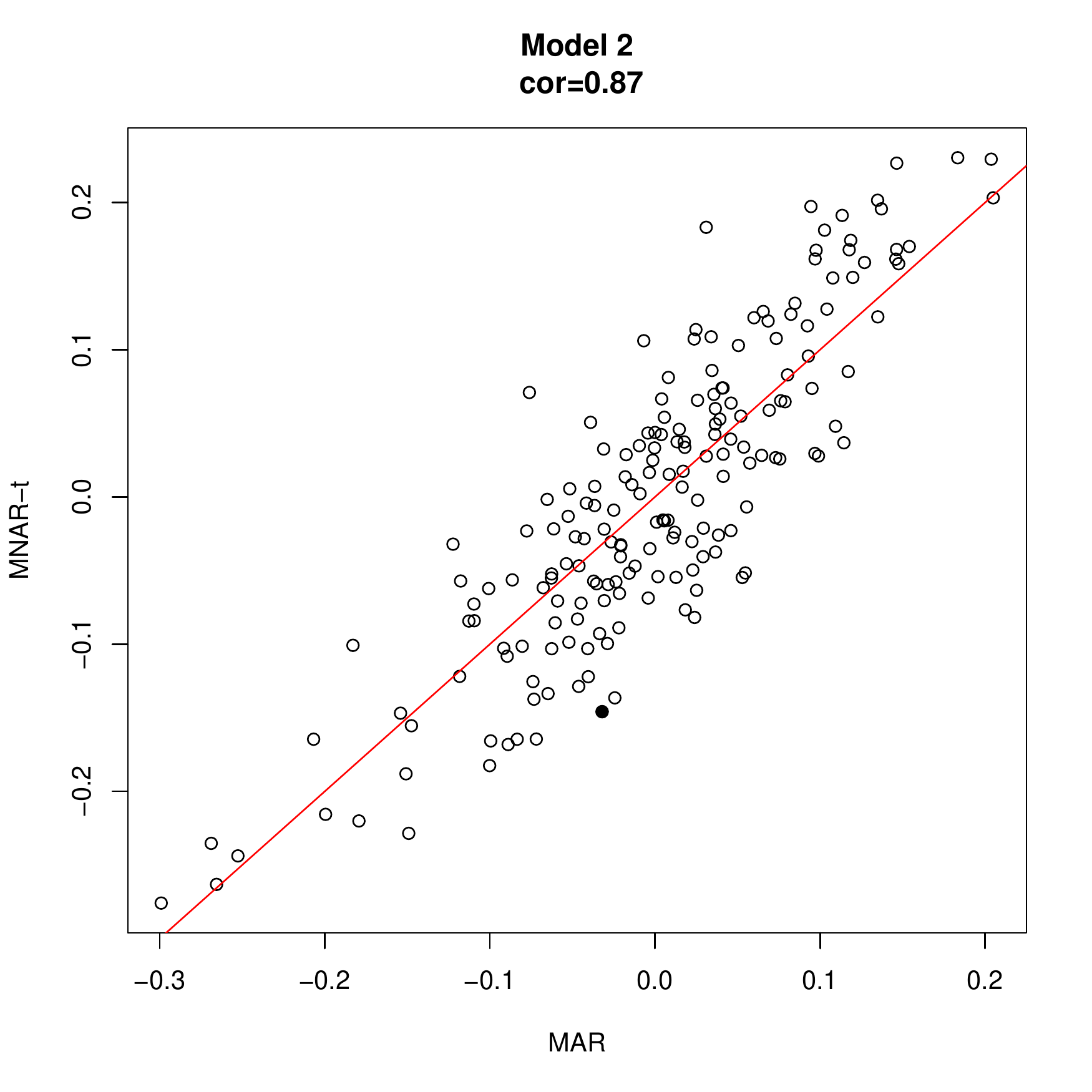}
\end{figure}

\citet{aaronson} rank teachers by the quartile of the relevant effect that their individual estimate falls in. While sometimes used in practice for personnel decisions, a simple division of the classrooms into quartiles does not account for the error in the estimates
of the classroom effects.
Analyzing the calculus data with MNAR-t leads to different classifications with the quartile method than those produced by MAR model. Thus, a teacher may receive a different evaluation based on the model assumed (either tacitly or explicitly) for the attendance mechanism. Using the method of \citet{aaronson}, some teachers move two (or even three) quartiles when evaluated with MNAR-t, as shown in Table \ref{table:quartile}. Figure \ref{plot:asu_joint_mar_2} plots the calculus 2 future year teacher effects from MNAR-t against the future effects from the MAR model. The quartile ranking appears to be relatively sensitive to the assumed nature of the missing data, although the confidence intervals for estimated teacher effects may also be wide.  Out of the 83 classrooms that change quartiles, 73 of those change only one quartile. These changes could be as simple as, for example, a shift from the 26th to the 24th percentile. 

\begin{table}
\centering
\caption{Quartiles of Calculus 2 Future Year Teacher Effects from MNAR-t (top) vs. MAR (left)}
\label{table:quartile}
\begin{tabular}{cccccc}\toprule
Quartile&1 & 2 & 3 & 4\\
\midrule\\
  1& 34 &11  &2 & 1\\
  2 &11 &20& 14&  2\\
  3 & 3 &14 &19 &11\\
  4 & 0 & 2 &12& 33
\end{tabular}
\end{table}

By contrast, \citet{lock07}, considering precision as well as ranking, only declare teacher effects as below/above average if their $90\%$ confidence (posterior credible) intervals are strictly below/above 0. The difference between MAR and MNAR-t is not as strong using this approach (see Table \ref{table:CI}), but some teachers still change categories under this more stringent criterion.

\begin{table}
\centering
\caption{$90\%$ Confidence Interval Rankings for Calculus 2 Future Year Teacher Effects from MNAR-t (top) vs. MAR (left)}
\label{table:CI}
\begin{tabular}{ccccc}\toprule
&- & 0 & +\\
\midrule
  -&5  &2 	 &0\\
  0 &2 &171 & 7\\
  + & 0 &0 &2
\end{tabular}
\end{table}

Following the suggestion of \citet{molen08}, we compare the fit of MNAR-t to that of MAR to see which classroom effects are most affected by the joint modeling of the attendance mechanism. The large amount of missing data in certain calculus classrooms means that the effects of those classrooms are attenuated toward zero due to the shrinkage properties of EBLUPs. This shrinkage property is normally desirable in VAMs, but in the case of potentially nonignorable dropout, we may lose information. For illustration we examine the records of one of the teachers most greatly down-weighted by MNAR-t in Figure~\ref{plot:asu_joint_mar_2}. This teacher's effect changed from $-0.03$ under MAR to $-0.14$ under MNAR-t in Model 1, and is represented by the solid circle in Figure~\ref{plot:asu_joint_mar_2}. Only $20\%$ of the students from this classroom completed calculus 3 (most of them failed the calculus 2 course), and those that did all received below-average grades in their respective calculus 3 classrooms. The calculus 2 teacher's effect on calculus 3 in the MAR model is less than 0, but is severely shrunk because only a few observations are present. It is possible that the poor performance of this teacher's students was due entirely to student attributes that were not included in the model: motivation, major, time of course during day, etc. However, this example illustrates how exploring the sensitivity of effects to the  attendance mechanism can lead to different conclusions about teachers.

The correlation matrix for the effects of calculus 2 teachers from Model 1 under MNAR-t appears in Figure \ref{fig:asu.cor}. The last column of these matrices, ``3 completion'', yields information about the correlation of the attendance effect of the calculus 2 teachers. A larger attendance effect means that relatively more of a teacher's students go on to complete calculus 3.  This effect is positively correlated with both the ``2 on 2'' effect and the ``2 on 3'' effect, so that the attendance effect is correlated with high grades of the teacher's students in both calculus 2 and calculus 3. However, the current and future year effects for calculus 2 teachers are not correlated. For this data set, observing that a teacher gives above- or below-average grades yields no information about how well the students of that teacher perform in calculus 3. Applications of VAMs to standardized test score data in the elementary school setting usually show a strong positive correlation between the current and future teacher effects \citep{mariano10,karlem}.

\begin{table}[htbp]
  \centering
\caption{Sensitivity analysis for Model 1. Standard errors are in parentheses.}
\label{tab:ASU}
    \begin{tabular}{@{}rrrrr@{}}\toprule
    &MAR&{MNAR-t}&MNAR-s&MNAR-b\\

    \midrule
  $\mu_1^y$ & -0.095 (0.027)  & -0.097 (0.028) & -0.092 (0.027) & -0.094 (0.028) \\
    $\mu_2^y$ & -0.154 (0.034)  & -0.161 (0.035) & -0.282 (0.035) & -0.284 (0.035) \\
    $\mu_2^r$ &       &  0.246 (0.026) & 0.307 (0.065) & 0.304 (0.041) \\
    $\sigma^2_1$  & 0.388 (0.023) & 0.385 (0.023) & 0.328 (0.020) & 0.330 (0.020) \\
    $\sigma^2_2$  & 0.292 (0.019)  & 0.293 (0.019) & 0.330 (0.019) & 0.329 (0.019) \\
    $\bsGamma_{stu} [1,1]$  & 0.618 (0.026)  & 0.620 (0.026) & 0.680 (0.026) & 0.674 (0.025) \\
    $\bsGamma_{stu} [2,1]$  &       &              & 0.637 (0.128) & 0.640 (0.065) \\
    $\bsGamma_{stu} [2,2]$  &       &              & 0.600 (0.633) & 0.610 (0.261) \\
   $\bsGamma_{1} [1,1]$  & 0.082 (0.015)  & 0.085 (0.015) & 0.077 (0.013) & 0.082 (0.015) \\
    $\bsGamma_{1} [2,1]$ & -0.004 (0.009)  & -0.001 (0.010) & -0.006 (0.009) & -0.002 (0.010) \\
   $\bsGamma_{1} [3,1]$  &       &  0.044 (0.015)      &       & 0.017 (0.013) \\
   $\bsGamma_{1} [2,2]$  & 0.028 (0.011)  & 0.031 (0.011) & 0.028 (0.010) & 0.030 (0.011) \\
   $\bsGamma_{1} [3,2]$  &              & 0.021 (0.009) &       & 0.010 (0.012) \\
   $\bsGamma_{1} [3,3]$  &       & 0.040 (0.014) &       & 0.052 (0.022) \\
   $\Gamma_{2}$ & 0.080 (0.015)  & 0.082 (0.015) & 0.082 (0.015) & 0.082 (0.015) \\
   $-2l$&&20022.7&19447.6&19436.7\\
   $\rho$&1&0.881&.994&.984\\
    \bottomrule
    \end{tabular}%
\end{table}%

\begin{figure}
\caption{Calculus 2 Future Year Effects: Model 1 MAR vs. Model 2 MAR.}
\label{plot:0v1}
\centering
\includegraphics[scale=.5]{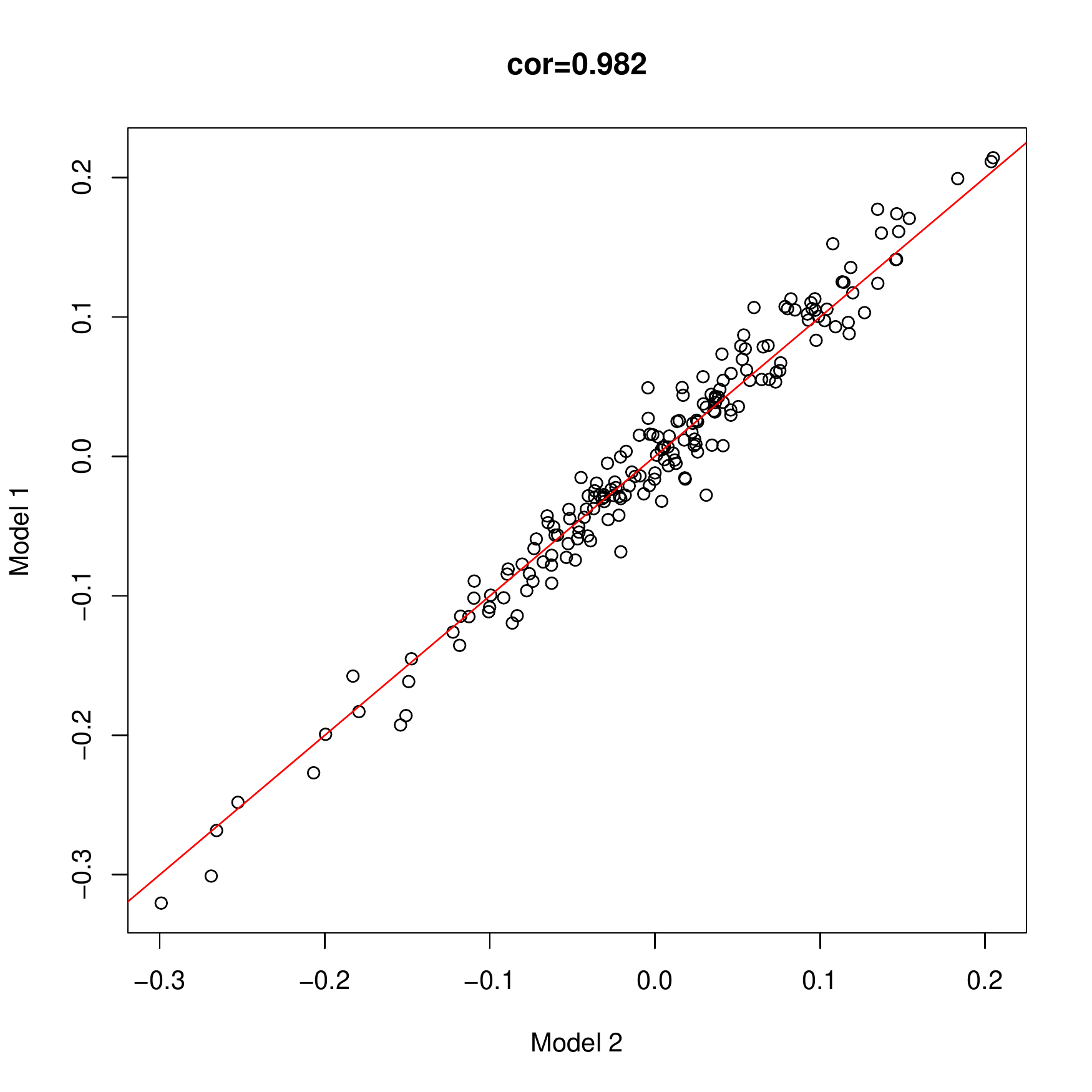}
\end{figure}

The correlations $\rho$ for Model 2 are nearly identical to those for Model 1 appearing in Table \ref{tab:ASU}. The correlations between MAR and MNAR-t, MNAR-b, and MNAR-s, for Model 2 are 0.870, 0.968, and 0.992, respectively. Furthermore, the fixed effects parameter estimates for Model 2 under MAR were nearly identical to those obtained under MNAR-t. The estimates appear in Table \ref{tab:model1}. Figure \ref{plot:0v1} compares the teacher ratings for Models 1 and 2 under an assumption of MAR. Interestingly, the addition of significant fixed effects to the model did not have a large impact on the EBLUPs. This contrasts with the difference seen between the rankings for Model 1 (and likewise Model 2) under MAR and MNAR-t seen in Figure \ref{plot:asu_joint_mar_2}.

\begin{table}[htbp]
  \centering
  \caption{Fixed effects estimates for Model 2 assuming MNAR-t. The estimates on the left are for the score model, while the estimates from the attendance model are on the right.}
    \begin{tabular}{rrrrrr}
    \addlinespace
    \toprule
          & $f(\bsy)$ &     &       & $f(\bsr)$ &  \\
    \midrule
		$\mu_1^y$& 0.602&(0.069)&&-&-\\
		$\mu_2^y$ and $\mu_2^r$& 0.539&(0.071)&&0.459&(0.096)\\
    Female &   -   &   -   &       & -     & - \\
    Male  & -0.155 & (0.035)  &       & 0.119  & (0.049) \\
    Asian &  -    & -     &       &  -    & - \\
    Black & -0.603 & (0.104)  &       & -0.315 & (0.147) \\
    Hispanic & -0.231 & (0.065)  &       & -0.203 & (0.094) \\
    Native Am. & -0.662 & (0.111)  &       & -0.375 & (0.156) \\
    Missing Race & 0.088  & (0.071)  &       & 0.126  & (0.106) \\
    White & -0.198 & (0.049)  &       & -0.199 & (0.072) \\
    SATQ-5 &   -  &   -   &       &  -    &-  \\
    SATQ-4 & -0.140 & (0.058)  &       & -0.023 & (0.086) \\
    SATQ-3 & -0.378 & (0.057)  &       & -0.053 & (0.084) \\
    SATQ-2 & -0.568 & (0.056)  &       & -0.234 & (0.081) \\
    SATQ-1 & -0.723 & (0.058)  &       & -0.255 & (0.083) \\
    Missing SATQ & -0.470 & (0.052)  &       & -0.182 & (0.076) \\
    \bottomrule
    \end{tabular}%
  \label{tab:model1}%
\end{table}%

\begin{figure}
 \caption{Correlation matrix of calculus 2 teacher effects from Model 1 under MNAR-t. ``2 on 2'' represents the effect of the calculus 2 teachers on calculus 2 grades, and ``2 on 3'' represents their effect on calculus 3 grades. ``3 completion'' gives the effect of calculus 2 teachers on calculus 3 attendance.}
 \label{fig:asu.cor}
\begin{align*}
cor(\bsGamma_1)=\bordermatrix{&\text{2 on 2 } &\text{2 on 3} &\text{3 completion}\cr
               \text{2 on 2}&1&-0.028&0.746\cr
                \text{2 on 3}& -0.028 &1& 0.596\cr
               \text{3 comp.}& 0.746&0.596&1}
\end{align*}
\end{figure}

\begin{figure}
\caption{Comparing the student score effects from MAR and MNAR-b under Model 2.}
\label{plot:student_score_effects}
\centering
\includegraphics[scale=.5]{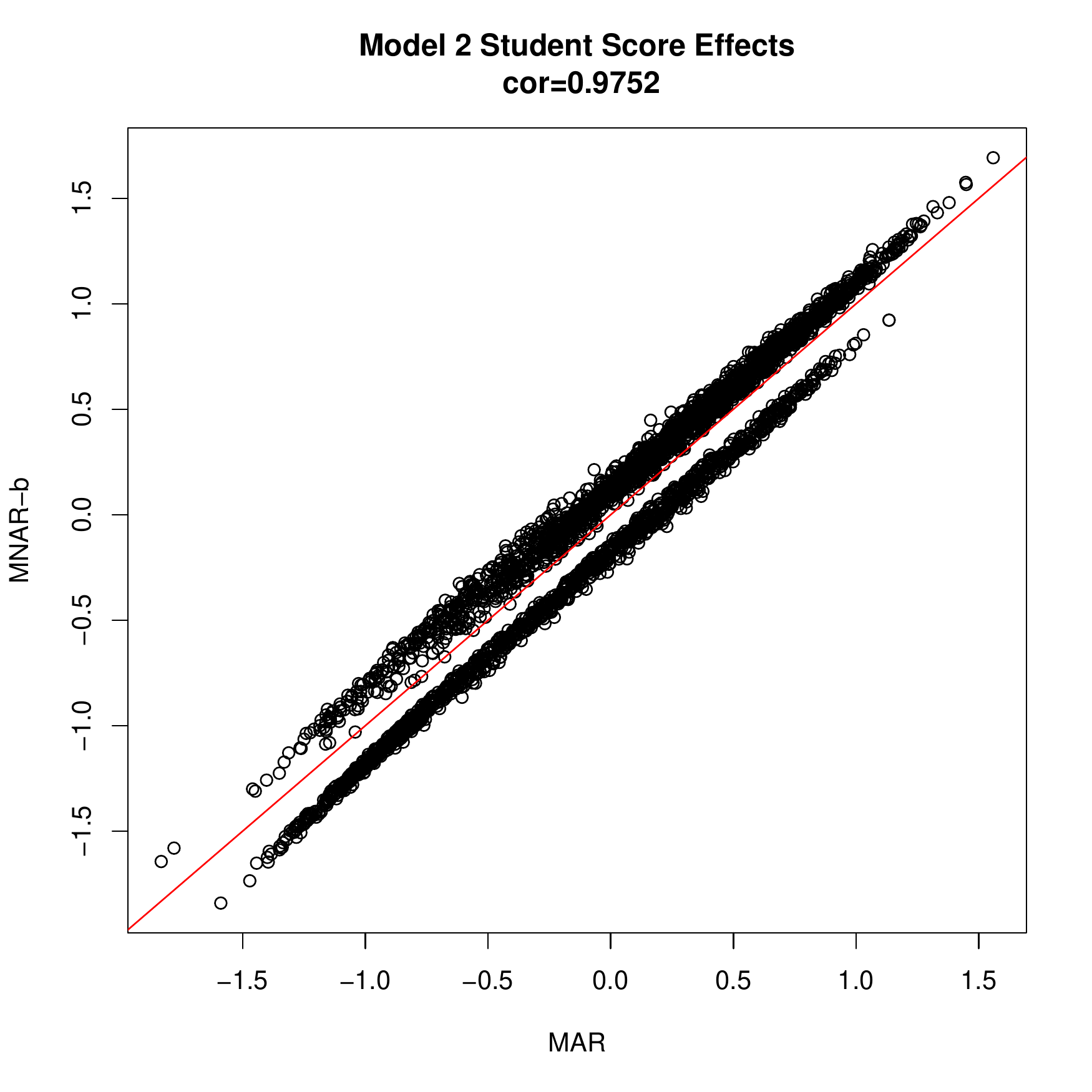}
\end{figure}

\begin{figure}
\caption{Comparing the student score and attendance effects from MNAR-b under Model 2.}
\label{plot:student_effects}
\centering
\includegraphics[scale=.5]{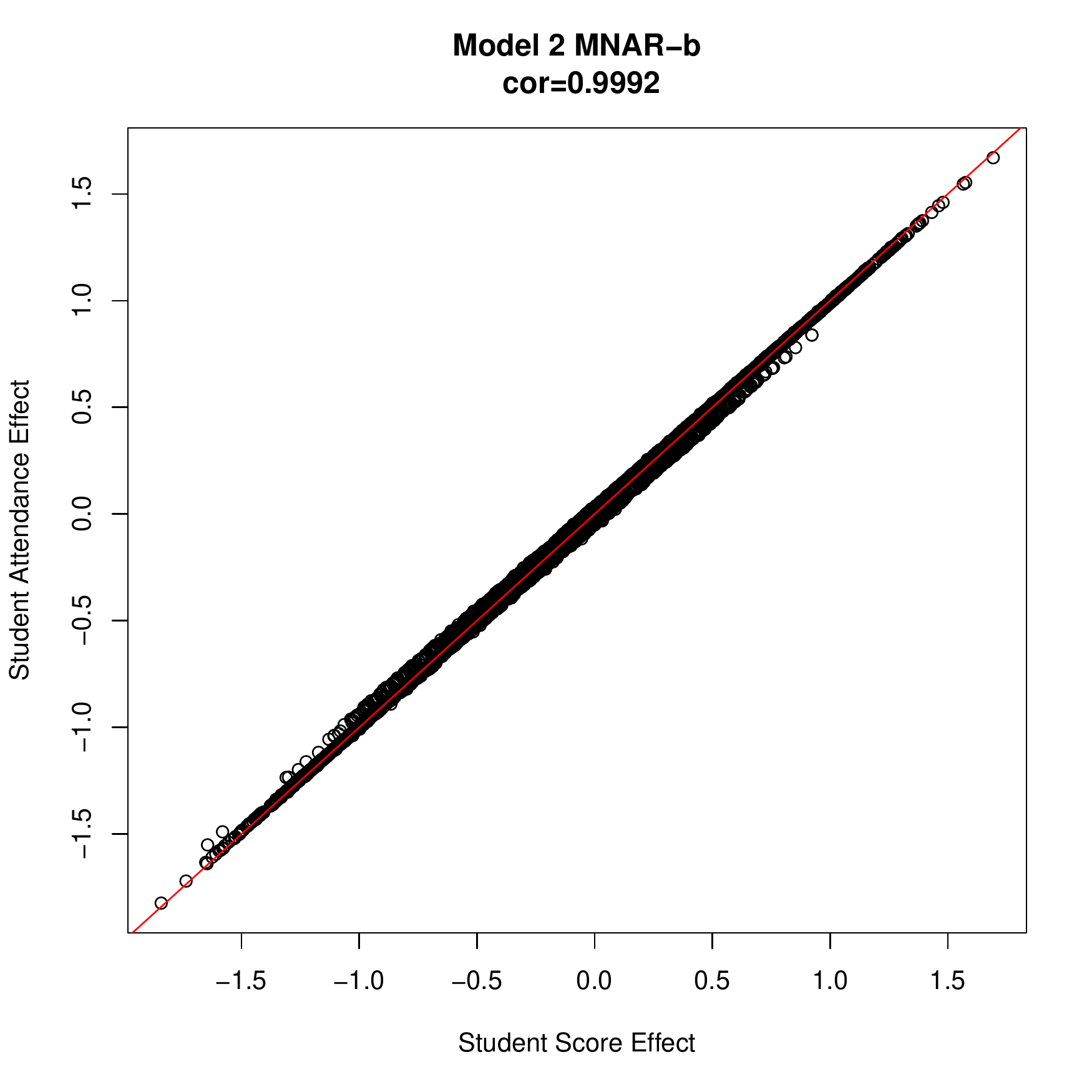}
\end{figure}

Figure \ref{plot:student_score_effects} compares the student score effects from Model 2 under assumptions MAR and MNAR-b (the results are nearly identical when comparing MAR and MNAR-s). Under MNAR-b, students who attended both years of calculus saw their score effect increase under MNAR-b, while those who attended only calculus 2 had their effects decreased. Figure \ref{plot:student_effects} shows the near-perfect correlation of student score and attendance effects in Model 2 under MNAR-b. Since there is only one year of observations (calculus 3) modeled by the attendance mechanism, the student attendance effects must borrow strength from the student score effects in order to be estimated. From Figure \ref{plot:student_effects}, it appears that these effects are identical. This is the same result we would have obtained for the student attendance effects if we had used a shared- rather than a correlated-parameter model. Under the correlated-parameter model, we would expect the correlation between these effects to decrease in situations where the attendance mechanism models more than a single year of observations. For the calculus example, the inclusion of student attendance effects under MNAR-b and MNAR-s requires an assumption that those effects will be identical to the student score effects.

The sensitivity analysis illustrates the influence that assumptions about the nature of missing data may have on the resulting teacher rankings. A challenge with MNAR models is that their fit for the missing data cannot be tested empirically. The fact that the likelihood for MNAR-b is larger than that of MNAR-t indicates that MNAR-b provides a better fit for the observed data ($\bsy^o$ , $\bsr$). It does not, however, indicate a better fit for the missing data $\bsy^m$. It is entirely possible that MNAR-t provides a better fit to $\bsy^m$ than MNAR-b: perhaps MNAR-b over-fits $(\bsy^o,\bsr)$. Without the ability to test the fit of the model to $\bsy^m$, the choice between MAR, MNAR-t, or any other relationship between the longitudinal and attendance processes requires an unverifiable assumption about the missing data process. It is interesting that the teacher effects in the score model are affected by the inclusion of teacher effects in the attendance model, but then return to their MAR values with the further inclusion of student effects in the attendance mechanism. This could represent a failure of the conditional independence (CI) assumption for the model MNAR-t \citep{mehm}. Nevertheless, the difference in teacher effects obtained between MAR and MNAR-t demonstrates how MAR estimates may be sensitive to some MNAR modifications while robust to others.

This is a non-standard application of a value-added model: typically, these models are applied to standardized test scores from elementary and secondary students, not to university data. Furthermore, inference usually focuses on the current year VAM effects. In this analysis, we focused on the future year effect from the GP VAM rather than the current year effects.  \cite{ballou}, \cite{lock07}, and \cite{mariano10} note that the effects from the first year included in the study are susceptible to bias due to non-random classroom assignment and capture the cumulative effects of prior teachers on those students. 

As with any observational data set, caution must be exercised when interpreting the results. Students were not randomly assigned to teachers, so effects ascribed to teachers may in fact be due to other factors. If students from majors that did not require calculus 3 tended to take calculus from certain instructors, then the attendance effects of those instructors would reflect the majors of their students rather than an impact of the teacher on taking calculus 3. We did not find evidence of clustering by major in the data set, but it is possible that time of day or other
confounding factors  may contribute to the estimated teacher effects.

\section{Elementary School Application}\label{sec:gradeschool}

We fit a different missing data mechanism to data from a large urban elementary school district. The data set tracks a cohort of 2834 students from grades 4 through
6, recording their score on a standardized math test each year.  The data set contains 102, 104, and 98 fourth, fifth, and sixth grade teachers,
respectively. Fixed effects representing the mean response in each year, race/ethnicity, and gender are included in both the score and the attendance mechanisms. In the elementary setting, students typically have no choice about whether to
progress to the next grade. In this setting, we would not expect grade $g$ teachers to have
an effect on whether their students take the test in grade $(g+1)$ but they might have
an effect on whether their students take the test in grade $g$. We therefore fit a different
model for the attendance process than for the university data. In this model, $\Lambda_{g[j]}$ represents the effect that the $j$-th grade $g$ teacher has on the probability of his or her students being measured in the same year $g$. A total of $421$ out of the $6657$ student observations with recorded teacher links are missing a test score.

Despite finding moderate correlations between the teacher effects in the score and attendance models (see Table~\ref{tab:chand} and Figure~\ref{chand6} for the parameter estimates), the estimates of teacher effects on scores are practically identical under each model adopted to explore the missing data mechanism. The correlations between teacher effects under MAR and MNAR-t are all greater than $0.992$; the plots are not displayed here because they are essentially straight lines. We also fit the model used in Section~3, exploring possible teacher effects on attendance in the following year, and likewise find that the model adopted for the missing data make little difference to the estimates of teacher effects on scores. This could be related to the fact that only around $6\%$ of the observations are missing. By contrast, around $40\%$ of the observations in the calculus example were missing.

\begin{table}[htbp]
  \centering
  \caption{Estimates from MNAR-t for elementary school data. The estimates on the left are for the score model, while the estimates from the attendance model are on the right.}
    \begin{tabular}{rrrrrr}
    \addlinespace
    \toprule
          & $f(\bsy)$ &     &       & $f(\bsr)$ &  \\
    \midrule
		$\mu_4^y$ and $\mu_4^r$& 24.303&(0.167)&&1.236&(0.097)\\
		$\mu_5^y$ and $\mu_5^r$& 25.289&(0.167)&&1.225&(0.094)\\
		$\mu_6^y$ and $\mu_6^r$& 26.315&(0.172)&&1.320&(0.099)\\
		$\sigma^2_4$& 1.489&(0.079)&&-&-\\
		$\sigma^2_5$& 1.028&(0.064)&&-&-\\
		$\sigma^2_6$&1.633&(0.081)&&-&-\\
		$\Gamma_{stu}$&3.899&(0.131)&&-&-\\
    Female &   -   &   -   &       & -     & - \\
    Male  & 0.039 & (0.082)  &       & 0.062  & (0.050) \\
    Asian &  1.500    & (0.226)     &       &  0.027    & (0.124) \\
    Black & - & -  &       & - & - \\
    Hispanic & 0.101 & (0.169)  &       & 0.346 & (0.092) \\
    Native Am. & 0.104 & (0.347)  &       & -0.190 & (0.173) \\
    White & 1.185 & (0.158)  &       & 0.356 & (0.086) \\
    \bottomrule
    \end{tabular}%
  \label{tab:chand}%
\end{table}%

\begin{figure}
\caption{Estimated blocks of the $\bsG$ matrix from MNAR-t. The covariance matrix is on the left, and the correlation matrix is on the right. Within each matrix, the current year score effects appear in the leftmost column, followed by future year score effects, and then by the current year attendance effect.}
\label{chand6}
$\bsGamma_4$:
\begin{flalign*}
&\begin{pmatrix}
0.648& 0.349& 0.332& 0.120\\
0.349& 0.225& 0.219& 0.099\\
0.332& 0.219& 0.238& 0.077\\
0.120& 0.099& 0.077& 0.099
\end{pmatrix}
\begin{pmatrix}
1.000& 0.914& 0.846& 0.474\\
0.914& 1.000& 0.947& 0.660\\
0.846& 0.947& 1.000& 0.498\\
0.474& 0.660& 0.498& 1.000
\end{pmatrix}&
\end{flalign*}
$\bsGamma_5$:
\begin{flalign*}
&\begin{pmatrix}
0.412& 0.165& 0.025\\
0.165& 0.084& 0.012\\
0.025& 0.012& 0.060
\end{pmatrix}
\begin{pmatrix}
1.000& 0.889& 0.157\\
0.889& 1.000& 0.165\\
0.157& 0.165& 1.000
\end{pmatrix}&
\end{flalign*}
$\bsGamma_6:$
\begin{flalign*}
&\begin{pmatrix}
0.441& 0.111\\
0.111& 0.112
\end{pmatrix}
\begin{pmatrix}
1.000&0.500\\
0.500&1.000
\end{pmatrix}&
\end{flalign*}
\end{figure}

In this data set from an elementary school district, the estimates of teacher effects on scores are insensitive to the choice of attendance mechanism (from those that were presented), though this does not imply that that the missing data mechanism is ignorable. This insensitivity may also be a function of the relatively small proportion of missing data in this example. \citet{graham} observes that all missing data are on a continuum between MAR and MNAR: we should focus on whether or not the likely violations of MAR matter to any practical extent. Even in such situations when the teacher effects do not show sensitivity to the choice of several different MNAR models, this class of correlated random effects models may still be useful for searching for abnormal features of the data set. For example, unlike in the university setting, we might not expect to see a strong relationship between current year teacher effects and next-year attendance effects. Yet some teachers might appear be outliers in bivariate plots of these effects, giving information about unusual cases in the data. As always, these potential outliers may be due to confounding factors, but they may indicate teachers with an unusual pattern.

\section{Summary}
\label{sec:summary}

We have developed a correlated random effects model to explore the sensitivity of teacher rankings from the GP VAM \citep{mariano10} to assumptions about the missing data process. In an application to calculus grades from a large university, the MAR teacher effects matched those obtained from two MNAR models that allowed the attendance process to depend on random student effects. The effects were robust even in the presence of significant correlation between random effects in the score and attendance models. If a given joint model is assumed to be correct, then correlation between the longitudinal and missingness processes indicates that the missing data are nonignorable. The finding highlights the point by \citet{graham} that the focus of a sensitivity analysis should not be on whether or not the MAR assumption has been violated, but rather on whether or not the violation is large enough to have practical implications.

The joint model MNAR-t, which allows for MNAR data under the specified attendance mechanism with included teacher effects, produces a different ranking and classification of the calculus 2 teacher future effects than the MAR GP model (the current year effects were unaffected). By contrast, MNAR-t produced roughly the same teacher rankings as the MAR model for the elementary school example. Likewise, \citet{mc10} did not find an appreciable difference in the results of their MNAR and MAR models while analyzing data from elementary school standardized scores, attributing the missingness to student characteristics. Three important differences between the calculus and the elementary school examples are the lack of standardization in the calculus grades, the larger percentage of missing data in the calculus example, and the greater potential for the calculus attendance trajectories of students to vary by teacher, due to the greater choice college students have in selecting future courses. In addition, the calculus rankings would have likely benefited from the inclusion of additional covariates such as the student major and the time of day that the course is offered. These factors may help explain the more profound changes to calculus teacher rankings resulting from the joint model MNAR-t.

In an application to elementary school data, none of the presented MNAR models produce a large number of significantly different teacher effects from those obtained under MAR. 
We would expect that in many elementary data settings, the teachers would have little effect on their
students' attendance at the test. However, the missing data models proposed in this paper could be used to identify unusual patterns in the data if such occurred.
In secondary school data, one might expect to see an effect of grade $g$ teachers on grade $(g+1)$ class taking, particularly with elective classes. For example, if the high schools require only two years of math, a sophomore math teacher may have an effect on his/her students' decisions to take advanced math classes. Thus, we would expect that the missing data models used for the calculus data in this paper would also be useful at the secondary level.

Value-added models are typically fit on observational data, not on data from a designed experiment. It is therefore always a possibility that the effects on student test scores that are ascribed to teachers are actually due to an unmeasured attribute of students who are assigned to that teacher. The same is true for the attendance models proposed in this paper. In the university setting, a teacher may have a low fraction of students proceed to calculus 3 if that teacher's students are in a discipline that does not require calculus 3. At the elementary school level, a teacher may be assigned a class with a large number of students who are exempt from the testing requirement, in which case the data are missing because of student rather than teacher characteristics. Thus, effects estimated for individual teachers must be interpreted carefully and other potential confounding factors need to be considered.

The methodology of this paper has been developed in the educational setting, but it applies in many other arenas as well. For example, longitudinal studies of medical interventions often have missing data, and the patients may be treated by several medical practitioners or hospitals. The methods of this paper can be used to evaluate effects of missing data in this context.

\section*{Acknowledgments}
This research was partially supported by the National Science Foundation under grant DRL-0909630, and by Arizona State University through a Dissertation Fellowship. Any opinions, findings, and conclusions or recommendations expressed in this material are those of the author and do not reflect the views of the National Science Foundation or Arizona State University. We would like to thank the three anonymous referees for their careful review and helpful suggestions that led to an improved paper.

\bibliographystyle{asabst}
\bibliography{disbib2}

\end{document}